\documentclass[aps,prb,showpacs,twocolumn,reprint,superscriptaddress]{revtex4-1}
\usepackage{amsmath}
\usepackage{amssymb}
\usepackage{graphicx}
\usepackage{color}
\usepackage{multirow}
\usepackage{dcolumn}
\usepackage{textcomp}
\usepackage{longtable}
\usepackage{makecell}
\usepackage{hyperref}
\hypersetup{hypertex=true, colorlinks=true, linkcolor=blue, anchorcolor=blue,citecolor=blue,urlcolor=blue}
\usepackage{bm}
\begin{document}
\title{Effects of substrate-surface reconstruction and orientation on spin-valley polarization in MoTe$_2$/EuO}

\author {Ao Zhang}
\author {Zisheng Gong}
\author {Ziming Zhu}
\email{zimingzhu@hunnu.edu.cn}
\affiliation{ School of Physics and Electronics, Hunan Normal University,
Key Laboratory for Matter Microstructure and Function of Hunan Province,
Key Laboratory of Low-Dimensional Quantum Structures and Quantum Control of Ministry of Education, Changsha 410081, China
}
\author {Anlian Pan}
\affiliation{Key Laboratory for Micro-Nano Physics and Technology of Hunan Province, 
College of Materials Science and Engineering, Hunan University, Changsha 410082, China
}
\author {Mingxing Chen}
\email{mxchen@hunnu.edu.cn}
\affiliation{ School of Physics and Electronics, Hunan Normal University,
Key Laboratory for Matter Microstructure and Function of Hunan Province,
Key Laboratory of Low-Dimensional Quantum Structures and Quantum Control of Ministry of Education, Changsha 410081, China
}

\begin{abstract}
We investigate the spin-valley polarization in MoTe$_2$ monolayer on (111) and (001) surfaces of ferromagnetic semiconductor EuO 
based on first-principles calculations.
We consider surface reconstructions for EuO(111).
We find that there is no direct chemical bonding between the reconstructed EuO(111) and the MoTe$_2$ overlayer,
in contrast to the case of the ideal EuO(111).
However, there is a strong hybridization between the states of MoTe$_2$ and the substrate states, 
which has a substantial impact on the valleys. 
The valley polarization due to the magnetic proximity effect is dependent on the detail of the interface structure,
which is in the range of a few meV to about 40 meV.
These values are at least one order of magnitude smaller than that induced by the ideal EuO(111). 
When the MoTe$_2$ monolayer is interfaced with EuO(001), the valley polarization is about 3.2 meV, 
insensitive to the interface structure.
By a low-energy effective Hamiltonian model, the effective Zeeman field induced by EuO(001) is about 27 T,
comparable to that for WSe$_2$/EuS obtained by experiment.
\end{abstract}


\maketitle
\section{INTRODUCTION}
Transition metal dichalcogenide (TMD) monolayers, MX$_2$ (M = Mo and W; X = S, Se, and Te), 
have engendered significant interest in two-dimensional materials for their unique structural and electronic properties
\cite{ref1nn400280c,ref2novoselov20162d,ref3manzeli20172d,ref6xu2014spin,
ref7wang2012electronics}.
They are direct band-gap semiconductors with both the valence band maximum (VBM) and the conduction band minimum (CBM) 
at the two inequivalent high symmetry points K and K$^\prime$.
Electrons in the two valleys can be used to encode information, 
thus making the TMD monolayers promising in valleytronics\cite{ref8coupledspin}.
An intriguing feature of the TMD family is the valley-dependent spin-momentum locking,
which results from the spin-orbit interaction along with the broken inversion symmetry\cite{ref8coupledspin,TMD_SOC_PRB_2011}. 
This spin-valley locking allows for the manipulation of the spin-valley polarization in the TMD monolayers,
which is of importance for spintronic and valleytronic applications.

Efforts have been devoted to the valley polarization in the TMD monolayers.
One approach is to optically pump the electrons in the valleys using circularly polarized light. 
This approach makes use of the inherent property of the TMD monolayers
that the right (left)-handed circularly polarized light are coupled with the K (K$^\prime$) valley
\cite{ref10Valley_Mak_2012,ref11Valley_Zeng_2012}. 
Another straightforward way is to break the time-reversal symmetry that gives rise to 
the degeneracy of the two valleys by means of external magnetic fields
\cite{ref12srivastava2015valley,ref13aivazian2015magnetic,ref14valleysplitting,ref15breaking}.
The valley polarization induced by the magnetic field is usually limited to 0.2 meV/T.
Recently, magnetic proximity effect has been proposed for manipulating the valley polarization.
For this purpose, the TMD monolayers are placed onto surfaces of magnetic semiconductors 
\cite{ref16Qi,ref17cheng,ref18mos2EuS,ref19ws2mno,ref20tmdmno,ref21ws2mno2,ref22mos2coo,ref23wte2fe3o4,
ref24wte2ymno3,ref25mose2crbr3,ref26wse2cri3,ref27wssemno,ref28typeIII,ref29wse2cri3,ref30Xie_2018}.
In particular, first-principles calculations predicted that
a MoTe$_2$ monolayer on EuO(111) has a valley polarization of over 300 meV\cite{ref16Qi,ref17cheng}. 
However, experiments found that the induced valley polarizations are less than 4 meV and less than 20 meV for WSe$_2$/EuS
\cite{ref31zhao2017enhanced} and WS$_2$/EuS\cite{ref32norden2019giant}, respectively.

The distinct difference in the valley polarization between the theoretical calculations and experiments may be associated with the difference 
in structural modeling of the heterostructures. 
Experimentally, the substrate-surface is most likely the (001) surface\cite{ref31zhao2017enhanced,ref32norden2019giant}. 
However, in the theoretical modelings, the (111) surface was used 
since it has the same type of lattice as the TMD monolayers and there is a small lattice mismatch between EuO(111) and MoTe$_2$
\cite{ref16Qi,ref17cheng}.
Moreover, substrate-surface reconstruction may be another factor for the difference in the valley polarization.
EuO and EuS are in the rocksalt structure, for which the (111) surface is polar. 
This type of surface, e.g., NaCl(111), MgO(111), NiO(111), tends to form various surface reconstructions 
\cite{ref33nacl,ref34nio2x2,ref35mgo,ref36cyclicmgo,ref37chargetransfermgo,ref38nio111,ref39rt3xrt3mgo,ref40theorymgo,ref41zhangmgo,ref42zhangnio,ref43mno}, 
which were unfortunately not considered in the previous studies.

In this paper, we investigate the effects of substrate-surface reconstruction and orientation on the valley polarization in MoTe$_2$/EuO 
by means of first-principles calculations.
We find that surface reconstruction plays an important role in determing the interface structure that
unlike the ideal EuO(111), there is no direct chemical bonding between the overlayer and the reconstructed surfaces.
The valley polarization is strongly dependent on the details of the interface structure for MoTe$_2$/EuO(111),
which varies from less than 1 meV to about 40 meV.
While for MoTe$_2$/EuO(001), it is about 3 meV for all the considered configurations of the interface. 

\section{COMPUTATIONAL DETAILS}
Our density-functional theory (DFT) calculations were performed using the Vienna \textit{ab initio} simulation package\cite{ref44vasp,ref45vasp1}. 
We use the projector augmented wave method to construct pseudopotentials\cite{ref46paw,ref47paw1}. 
The plane-wave energy cutoff is 480 eV. 
The generalized gradient approximation as parametrized by Perdew-Burke-Ernzerhof exchange-correlation functional\cite{ref48pbe} 
is adopted for the exchange-correlation functional. 
We use the dispersion-corrected DFT-D3 method to account for 
the influence of van der Waals (vdW) dispersion forces between the overlayer and the substrate\cite{ref49vdw}.
The EuO(111) surface substrate is modeled by a thirteen atomic layers slab, 
which is separated from its periodic images by $\sim$20 \AA{} vacuum regions.
We employ the DFT + U method for EuO to account for the strong correlation effects associated with the 4$f$ electrons.
We take the parameters for the Coulomb and exchange interaction, i.e., $U$ and $J$,
given by previous studies\cite{ref16Qi,ref17cheng},
that is, for Eu-4$f$ orbitals, $U$ and $J$, are 8.3 eV and 0.77 eV, respectively.
While for O-2$p$ orbitals, they are 4.6 eV and 1.2 eV, respectively. 
7 $\times$ 7 $\times$ 1 and 14 $\times$ 7 $\times$ 1 $\Gamma$-centered Monkhorst-Pack grids of $k$-points were used to sample the surface Brillouin zone for
MoTe$_2$/otopolar-EuO(111) and MoTe$_2$/P(1 $\times$ 2)-EuO(111), respectively. And a 13 $\times$ 3 $ \times$ 1 $k$-mesh was used for MoTe$_2$/EuO(001).
Structural relaxations were done with a threshold of 0.01 eV/\AA{} for the residual force on each atom.

\section{RESULTS AND DISCUSSIONS}
\subsection{Structural properties}
We begin by discussing the structural properties of the MoTe$_2$/EuO(111) interface.
Unlike the bulk phase, the ideal (111) surface of EuO is metallic as revealed by previous studies\cite{ref16Qi,ref17cheng}.
The metallic behavior is caused by the dangling bonds at the surface, 
suggesting a strong tendency of reconstruction for this type of surface.
We consider two types of the surface reconstruction for EuO(111), the octopolar reconstruction and the P(1 $\times$ 2) reconstruction.
These reconstructions have been revealed for the (111) surface of NaCl, MgO, NiO and many other polar rocksalt surfaces\cite{ref33nacl,ref34nio2x2,ref35mgo,ref36cyclicmgo,
ref37chargetransfermgo,ref38nio111,ref39rt3xrt3mgo,ref40theorymgo,ref41zhangmgo,ref42zhangnio,ref43mno}.

\begin{figure}[htbp]
\centering
\includegraphics[width=8.7cm]{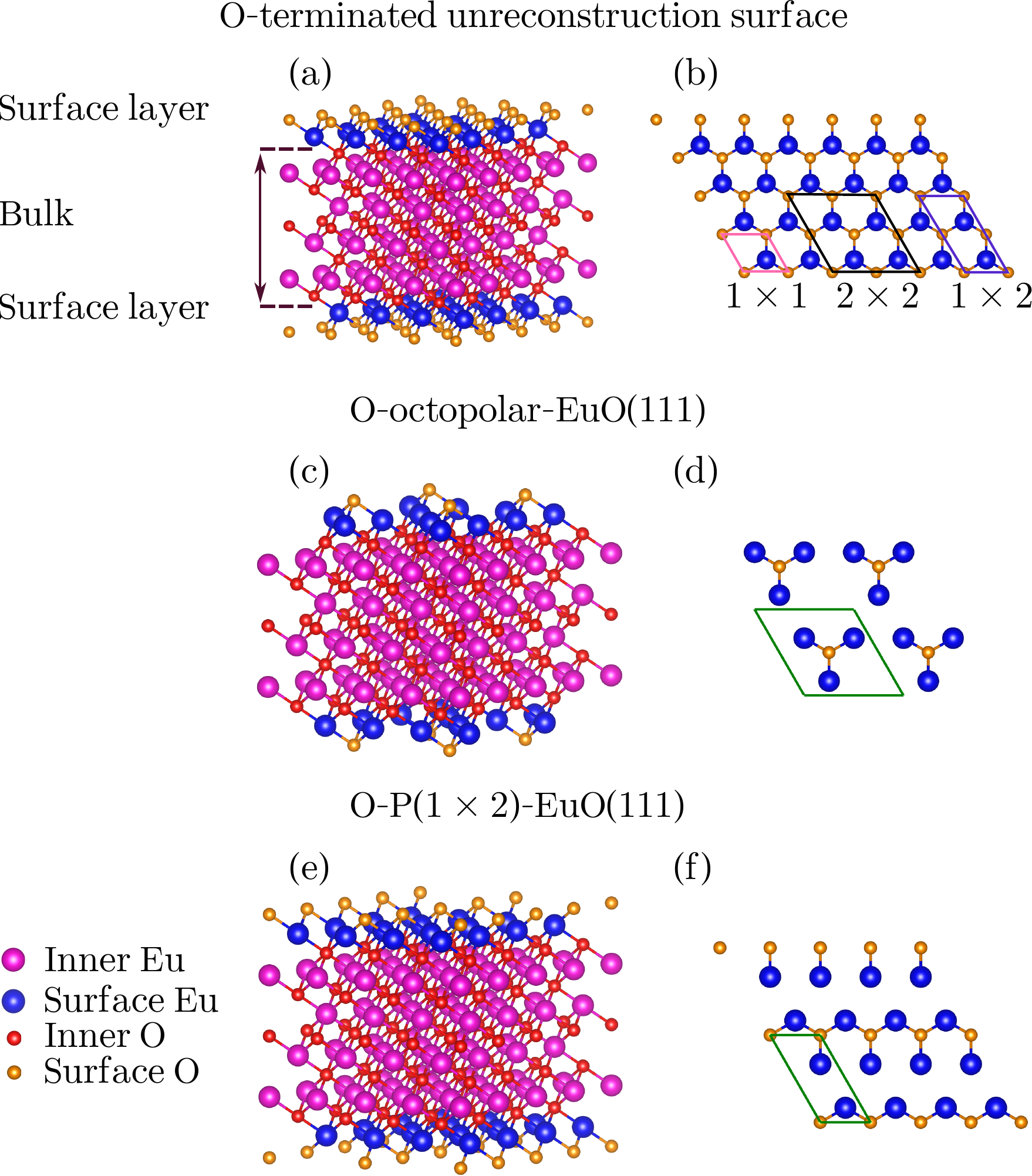}
\caption{Geometric structures for reconstructed EuO(111) with O-termination.
(a) Perspective view and (b) top view of the unreconstructed surface.
(c) Perspective and (d) top views of octopolar-reconstructed surface, i.e., octopolar-EuO(111).
(e) Perspective and (f) top views of P(1 $\times$ 2)-EuO(111).
In (b), (d), and (f) only the atoms in the surface layer are shown.
In (b), the pink, black, and purple boxes represent the $1 \times 1$ primitive cell (unreconstructed surface),
the 2 $\times$ 2, and 1 $\times$ 2 supercells, respectively.
The 2 $\times$ 2 supercell is used to build the octopolar-reconstructed surface shown in (c) and (d).
The 1 $\times$ 2 supercell is used to build the P(1 $\times$ 2) reconstruction shown in (e) and (f).
The green boxes denote the unit cell for the reconstructed surfaces.
Eu-terminated surfaces have similar structures with the O-terminated ones.}
\label{fig1}
\end{figure}

Figure~\ref{fig1} shows the structural models for the two types of surface reconstructions for O-terminated EuO(111).
The octopolar reconstruction shown in Figs.~\ref{fig1}(c, d) is derived from a 2 $\times$ 2 surpercell of the ideal/unreconstructed EuO(111).
Specifically, it is obtained by removing 75\% O atoms from the top layer of the surface and 
removing 25\% Eu atoms from the Eu layer underneath the surface oxygen layer.
The P(1 $\times$ 2) reconstruction (see Figs.~\ref{fig1}(e) and (f)) is built by removing every other oxygen atoms from the surface layer 
in a 1 $\times$ 2 supercell of the ideal EuO(111).
Likewise, the Eu-terminated octopolar structure is obtained by removing 75\% surface Eu atoms and 25\% O atoms underneath the surface Eu layer.
While in the Eu-terminated P(1 $\times$ 2) structure, we remove every other Eu surface atoms in the surface layer of 
a 1 $\times$ 2 supercell of EuO(111). 
Hereafter, Eu-octopolar-EuO(111) and O-octopolar-EuO(111) refer to the octopolar-reconstructed EuO(111) with Eu- and O-termination, respectively.
While Eu-P(1 $\times$ 2)-EuO(111) and O-P(1 $\times$ 2)-EuO(111) stand for the P(1 $\times$ 2)-reconstructed EuO(111) with Eu- and O-termination, respectively.
In our calculations, we consider the surface reconstructions for both sides of the slabs.
As a result, we find that Eu-octopolar-EuO(111), O-octopolar-EuO(111) and O-P(1 $\times$ 2)-EuO(111) are semiconducting,
except for that the Eu-P(1 $\times$ 2)-EuO(111) shows a half-metal behavior.
The calculated band gaps of Eu-octopolar-EuO(111), O-octopolar-EuO(111), and O-P(1 $\times$ 2)-EuO(111) 
are 0.69 eV, 0.80 eV, and 0.58 eV, respectively. 
Thus, one may expect that these reconstructed surfaces are more stable than the ideal EuO(111).

\begin{figure}[htbp]
\centering
\includegraphics[width=8.7cm]{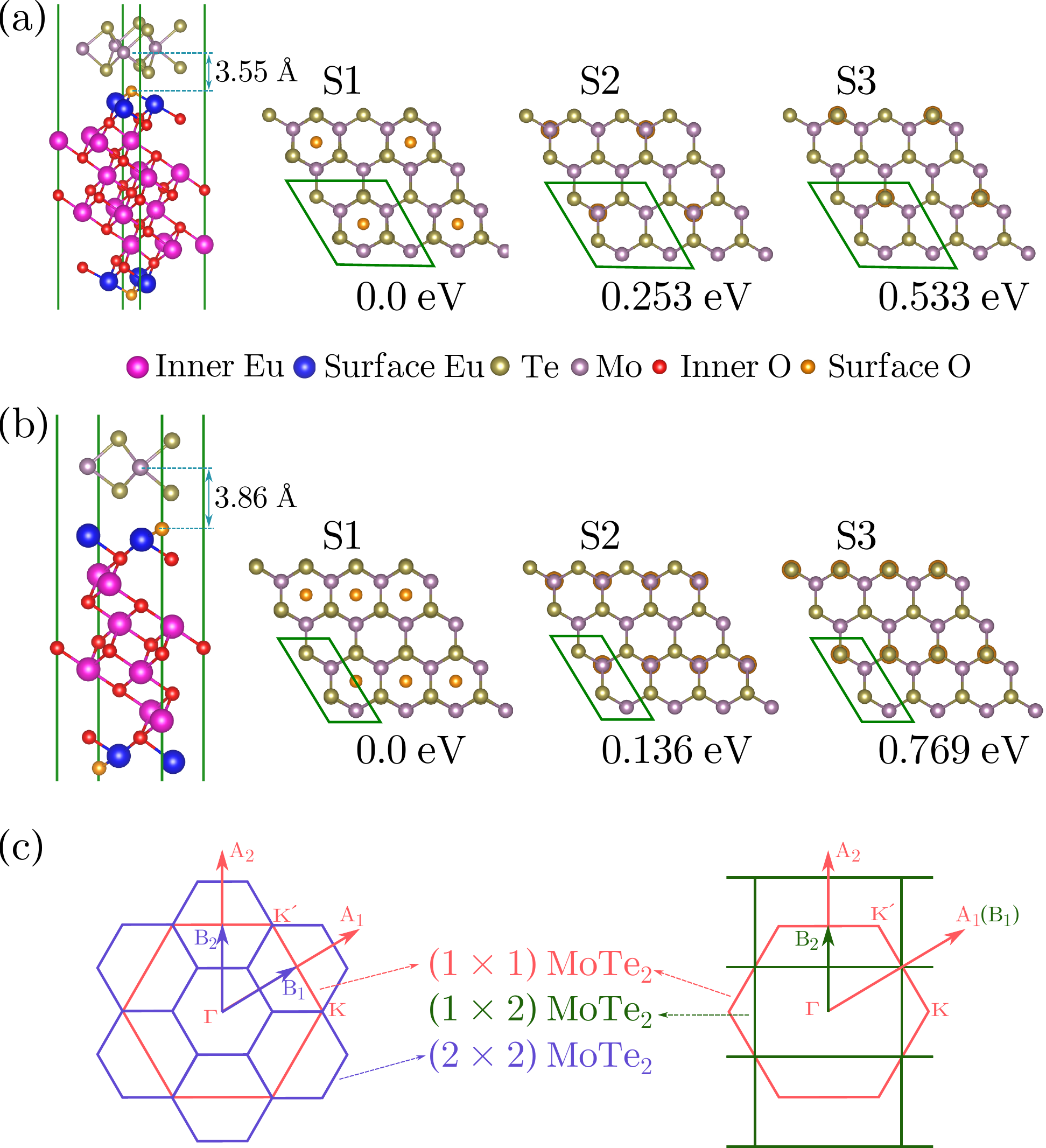}
\caption{Geometric structures of MoTe$_2$/EuO(111).
(a) Perspective view and top view of a 2 $\times$ 2 supercell of MoTe$_2$ on O-octopolar-EuO(111).
(b) Perspective view and top view of a 1 $\times$ 2 MoTe$_2$ on O-P(1 $\times$ 2)-EuO(111).
Three configurations are considered for each type of interface.
The total energy of each configuration relative to that of S1 is shown below the structure.
The green boxes in (a) and (b) denote the unit cell for MoTe$_2$/EuO(111).
(c) Brillouin zones of the (1 $\times$ 1) (red), (1 $\times$ 2) (green), and (2 $\times$ 2) (purple) supercells for MoTe$_2$, respectively.
\textbf{\textit{A$_i$}} and \textbf{\textit{B$_i$}} ($i$ = 1, 2) denote the reciprocal lattice vectors of the (1 $\times$ 1) primitive cell and the
2 $\times$ 2 (1 $\times$ 2) supercell.}
\label{fig2}
\end{figure}

We now turn to the MoTe$_2$/EuO(111) heterostructures, 
for which the geometric models are shown in Fig.~\ref{fig2}.
There are two types of heterostructures/interfaces if we classify them according to the surface reconstruction of EuO(111),
i.e., MoTe$_2$/octopolar-EuO(111) (Fig.~\ref{fig2}(a)) and MoTe$_2$/P(1 $\times$ 2)-EuO(111) (Fig.~\ref{fig2}(b)), respectively. 
Since there is a small lattice mismatch (less than 3\%) between the MoTe$_2$ monolayer and the ideal EuO(111),
we have the same number of unit cells of MoTe$_2$ and the unreconstructed EuO(111) in the slab models for the heterostructures. 
Specifically, there is a 2 $\times$ 2 supercell of MoTe$_2$ in the unit cell of MoTe$_2$/octopolar-EuO(111).
Likewise, there is a 1 $\times$ 2 supercell of MoTe$_2$ in the unit cell of MoTe$_2$/P(1 $\times$ 2)-EuO(111).
For each type of MoTe$_2$/EuO(111) heterostructures, 
we have considered three types of stacking between the MoTe$_2$ monolayer and EuO(111) substrate 
for both the O- and Eu-terminated surfaces.
These configurations are referred to as S1, S2, and S3, respectively.
As an example, we show in Fig.~\ref{fig2}(a) the detail of the interface structures of MoTe$_2$/O-terminated-EuO(111).
In S1, the hollow site of the MoTe$_2$ monolayer is on the top of the surface O atom.
While in S2 (S3), the surface O atom is right below Mo (Te).

The overlayer well preserves its geometric profile upon structural relaxation.
And there is no direct chemical bonding between the MoTe$_2$ monolayer and the substrate.
The layer distances between Mo and the surface outmost atom
for MoTe$_2$/O-octopolar-EuO(111) are 3.55 \AA, 4.07 \AA, and 4.06 \AA\ for S1, S2, and S3, respectively.
While for MoTe$_2$/O-P(1 $\times$ 2)-EuO(111), they are about 3.86 \AA, 4.26 \AA, 4.41 \AA\ for S1, S2, and S3, 
respectively. S1 is found to be more stable than the other two configurations for both types of heterostructures. 
It is 0.253 eV and 0.533 eV lower than S2 and S3 for the octopolar MoTe$_2$/EuO(111), respectively.
For the MoTe$_2$/O-P(1 $\times$ 2)-EuO(111), the energy difference between S1 and S2 (S3) is 0.136 eV (0.769 eV).

\subsection{Electronic structure of MoTe$_2$/EuO(111)}
With the geometric structures, we now proceed to the electronic structure of the MoTe$_2$/EuO(111) heterostructures.
Note that there are supercells of MoTe$_2$ in the structural models of the heterostructures, which cause band foldings.
To eliminate the folded bands, a band unfolding procedure is necessary. 
For interfaces like the systems in the present study, 
we use the layer $k$-projection method as implemented in program KPROJ\cite{ref50KPROJ}.
In this way, the unfolded bands were obtained by projecting the supercell wavefunctions of the MoTe$_2$ monolayer onto 
the $k$-points in the Brillouin zone (BZ) of the primitive cell.
Fig.~\ref{fig2} (c) shows the BZs for the primitive cell and supercells of the MoTe$_2$ monolayer and corresponding high symmetry points.

\begin{figure*}[htbp]
\includegraphics[width=0.95\textwidth]{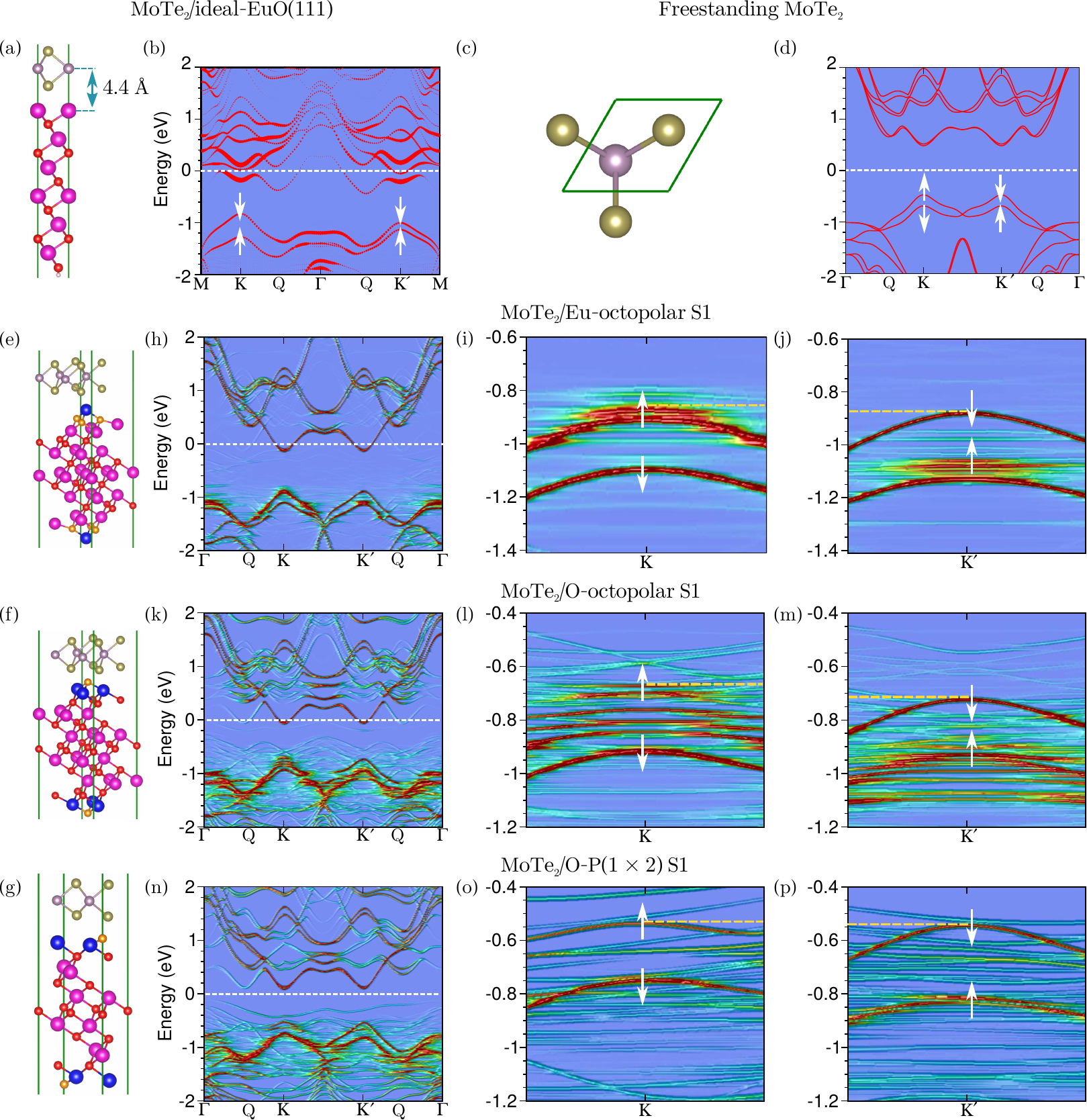}
\caption{Unfolded band structures for MoTe$_2$/EuO(111). 
(a, c) The geometric structures of MoTe$_2$/ideal-EuO(111) and the freestanding MoTe$_2$ monolayer, respectively.
(b) and (d) The band structures of MoTe$_2$ corresponding to (a) and (c), respectively. (e)-(g) The lowest-energy 
structure for MoTe$_2$ on Eu-octopolar, O-octopolar, and O-terminated P(1 $\times$ 2) surfaces of EuO(111), respectively. (h) Unfolded 
bands for MoTe$_2$ for the structure shown in (e). (i) and (j) unfolded bands around K and K$^\prime$, respectively. (k)-(m) and (n)-(p) 
Corresponding plots for the structures shown in (f) and (g), respectively. The white dashed line in (b), (d), (h), (k), and (n) represents 
the Fermi level. The yellow dashed line marks the top of the valley. 
The white arrows denote the spin components of the bands.}
\label{fig3}
\end{figure*}

Figure~\ref{fig3} shows the unfolded/$k$-projected bands for the MoTe$_2$ overlayer in MoTe$_2$/EuO(111). 
We only show the band structure for the lowest-energy configuration for each type of heterostructure,
namely, the S1 configurations shown in Fig.~\ref{fig2}. 
We have not attempted calculations of MoTe$_2$/EuO(111) for the substrate with Eu-terminated P(1 $\times$ 2) reconstruction
since this type of surface reconstruction is energetically higher than others.
The results for the freestanding MoTe$_2$ and MoTe$_2$/ideal-EuO(111) are shown for comparison (Figs.~\ref{fig3}(a)-(d)),
which are in good agreement with previous studies\cite{ref16Qi,ref17cheng}.
Figs.~\ref{fig3}(e)-(j) and (f)-(n) are for MoTe$_2$/Eu-octopolar-EuO(111) and MoTe$_2$/O-octopolar-EuO(111), respectively. 
While Figs.~\ref{fig3}(g)-(p) are for MoTe$_2$/O-P(1 $\times$ 2)-EuO(111).

One can see from Figs.~\ref{fig3}(h), (k) and (n) that the conduction bands of MoTe$_2$ preserve well 
in the presence of the substrate.
For MoTe$_2$/octopolar-EuO(111)(see Figs.~\ref{fig3}(e, f)), 
the Fermi levels move into the conduction band, similar to the heterostructure with ideal EuO(111)\cite{ref16Qi,ref17cheng} (see Fig.~\ref{fig3}(b)).
However, the Fermi levels of our systems are much closer to the gap than that of MoTe$_2$/unreconstructed-EuO(111).
One can see from Figs.~\ref{fig3}(h, k) that the Fermi level is about 0.1 eV above the CBM. 
Whereas, previous studies reveal that it is about 0.4 eV above the CBM for MoTe$_2$/unreconstructed-EuO(111)\cite{ref16Qi,ref17cheng}.
For MoTe$_2$/P(1 $\times$ 2)-EuO(111), the Fermi level even lies in the gap (see Fig.~\ref{fig3} (n)).
Another distinct effect of the surface reconstruction on the electronic structure of MoTe$_2$ is related to the nature of the band gap.
In the case of MoTe$_2$/ideal-EuO(111), the overlayer undergoes a direct-indirect band gap transition
 due to that the strong bonding between MoTe$_2$ and the substrate pushes the $Q$ point down to a lower energy than the CBM (Fig.~\ref{fig3}(b)).
In contrast, the direct band gap of the MoTe$_2$ monolayer is maintained upon interfacing with the reconstructed surfaces.

\begin{figure*}[htbp]
\centering
\includegraphics[width=0.95\textwidth]{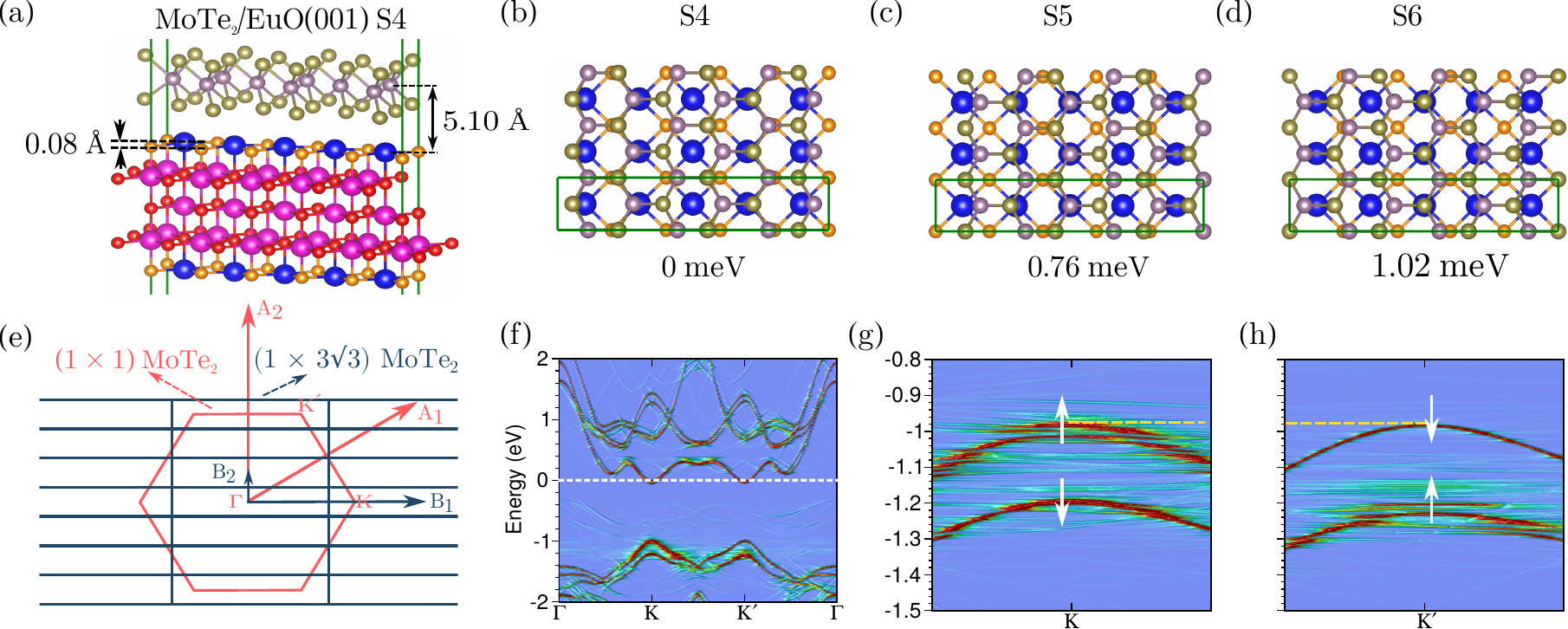}
\caption{Unfolded bands for MoTe$_2$/EuO(001).
(a) Geometric structures for MoTe$_2$/EuO(001).
The energy difference relative to the lowest-energy structure S4 is given below each configuration.
(e) Brillouin zones of the (1 $\times$ 1) and the 1 $\times$ 3$\sqrt{3}$ MoTe$_2$.
\textbf{\textit{A$_i$}} and \textbf{\textit{B$_i$}} ($i$ = 1, 2) denote the reciprocal lattice vectors of
the (1 $\times$ 1) primitive cell and the 1 $\times$ 3$\sqrt{3}$ supercell, respectively.
(f) Unfolded band structure for MoTe$_2$ and (g), (h) zoom-in band structure around K and K$^\prime$. 
The white arrows denote the spin orientation at the K and K$^\prime$ points.}
\label{fig4}
\end{figure*}

The reconstructed surfaces do have substantial impact on the valleys of the MoTe$_2$ monolayer.
In the two panels on the right side of Fig.~\ref{fig3}, we show the zoom-in unfolded bands near K and K$^\prime$.
One can see that the spin-up states of the MoTe$_2$ monolayer are modified by the interaction with the substrate states.
The reason is that there is a strong spin-dependent hybridization between the states of MoTe$_2$ and the substrate
since near the K and K valleys there are mainly the majority (spin-up) states of the substrate (not shown).
For MoTe$_2$/O-octopolar-EuO(111), the strong hybridization drives the spin-up band into multiple subbands, 
so that it is difficult to identify the VBM at K.
In this case, we define the state to which the contribution of the MoTe$_2$ monolayer over 50\% to be the VBM of it.
The spin-dependent interaction breaks the time-reversal symmetry that gives rise to the degeneracy of the electronic bands at K and K$^\prime$,
thus leading to the so called valley polarization. We define the valley polarization as $\Delta_{\mathrm{valley}}^{v, \tau} \equiv 
E_{\uparrow}^{v, \tau}-E_{\downarrow}^{v,-\tau}$. Here $v$ denotes the valence band, $\tau$ represents the valley index.
The valley polarizations in our systems are dependent on the surface reconstruction of the substrate and the specific stacking between them.
They are about 7.56 meV and 25.62 meV for MoTe$_2$/Eu-octopolar-EuO(111) and MoTe$_2$/O-octopolar-EuO(111), respectively.
We have also examined the valley polarization for a different stacking, e.g., S2 in Fig~\ref{fig2}(b), 
for which the valley polarization is about 32.77 meV.
For MoTe$_2$/O-P(1 $\times$ 2)-EuO(111), the valley polarization is about 12.07 meV.
These valley polarizations are much smaller than that for MoTe$_2$ on the ideal EuO(111),
for which it is about 300 meV as reported by previous studies\cite{ref16Qi,ref17cheng}.
These differences indicate that substrate-surface reconstruction plays an important role 
in determing the electronic structure of the TMD monolayers.

\subsection{Valley polarization in MoTe$_2$/EuO(001)}
We now discuss the valley polarization in MoTe$_2$/EuO(001). 
We use a slab consists of five-atomic layers to model EuO(001). 
One can see from Fig.\ref{fig4}(a) that this surface is flat since the Eu and O atoms almost have the same $z$ coordinate.
Unlike the ideal EuO(111), our DFT calculation of the slab find that EuO(001) is semiconducting.
Therefore, the surface reconstruction is not considered in our calculations of MoTe$_2$/EuO(001).
Note that EuO(001) and the MoTe$_2$ monolayer have different types of lattice,
which requires a large supercell to model the heterostructure.
In our model, the MoTe$_2$ monolayer is in a 1 $\times$ $3\sqrt{3}$ supercell and 
EuO(001) is in a 1 $\times$ 5 supercell.
As a result, the lattice mismatch between them is about 1.91\% for the \bm{$A_1$} direction and -1.94\% for the \bm{$A_2$} direction.
We have considered three different configurations between the MoTe$_2$ monolayer and EuO(001), 
which are denoted as S4, S5, and S6, respectively.
The geometric structures are shown in Figs.~\ref{fig4}(b)-(d).
S5 and S6 are obtained by translating the overlayer in S4 by 2.02 \AA, 4.10 \AA, respectively.
The layer distances for these configurations are about 5.10 \AA, 5.07 \AA, and 5.09 \AA{} for S4, S5, and S6, respectively.
These configurations show pretty much similar band structures.
Fig.~\ref{fig4}(f) depicts the unfolded structure for the MoTe$_2$ monolayer in S4.
The Fermi level of this type of heterostructure is closer to the CBM than those for MoTe$_2$/Octopolar-EuO(111).
The valley polarization is about 3.17 meV.
This value is comparable to that for WSe$_2$/EuS\cite{ref31zhao2017enhanced}, 
where the substrate-surface is believed to be the (001) surface.
Therefore, the magnetic proximity effect induced by magnetic substrates is surface-orientation dependent. 

\subsection{Band alignment}
\begin{figure}[htbp]
\centering
\includegraphics[width=8.7cm]{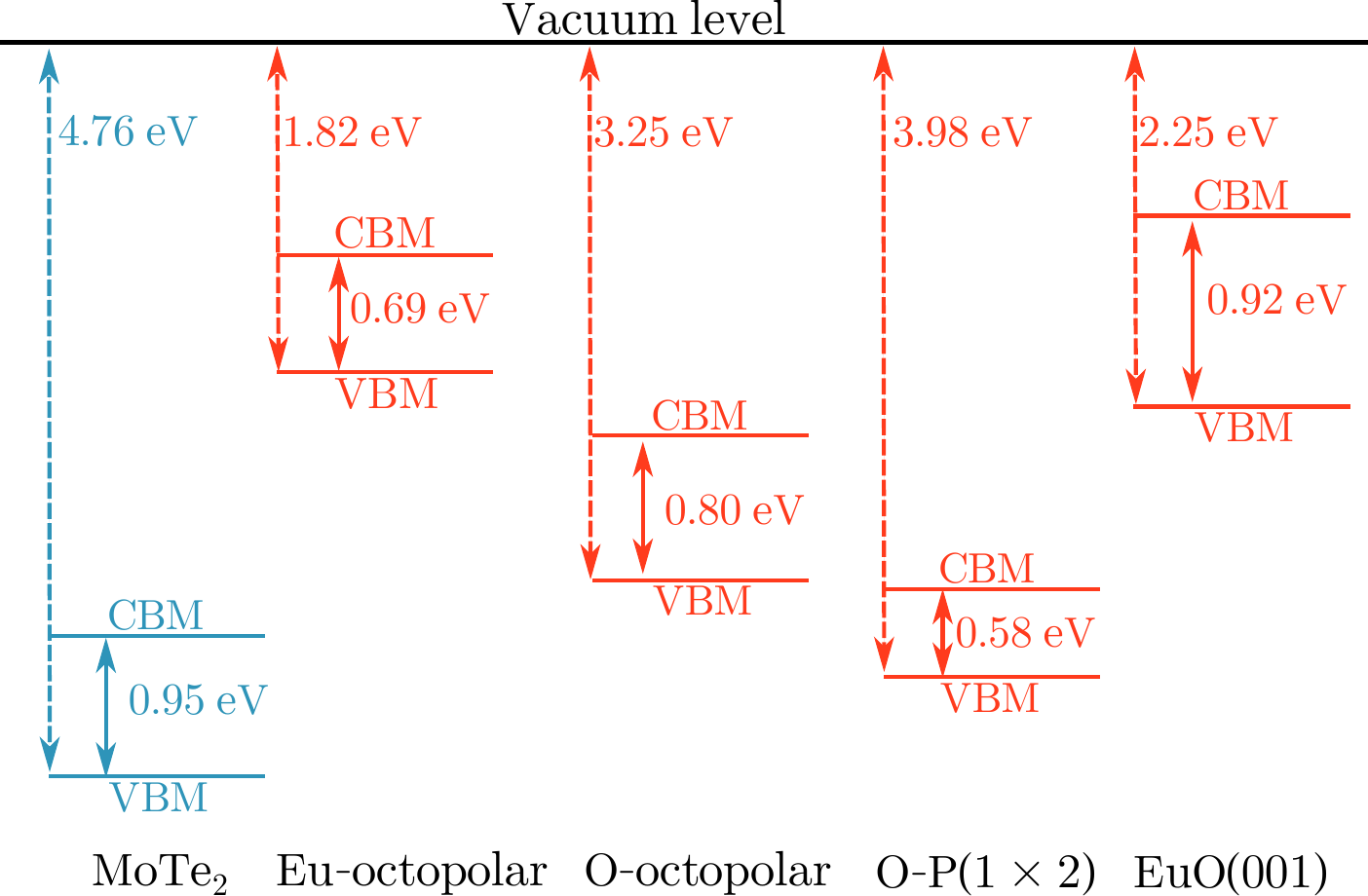}
\caption{Band alignments of MoTe$_2$ and EuO surfaces.
VBM and CBM denote the valence band maximum and the conduction band minimum, respectively. Work functions and the sizes of the band gaps are given.}
\label{fig5}
\end{figure}

Figure~\ref{fig5} shows the work functions of the isolated systems and 
the band alignments between the MoTe$_2$ monolayer and various EuO surfaces.
One can see that there is a type-II band alignment between the  monolayer and O-P(1 $\times$ 2)-EuO(111),
which gives rise to the location of the Fermi level right in the band gap.
However, the band alignments between MoTe$_2$ and the other surfaces are of type-III,
that is, the CBM of the MoTe$_2$ is energetically lower than the VBM of EuO surfaces.
Thus, the Fermi levels of their heterostructures move into the conduction band of the MoTe$_2$ monolayer.

\subsection{Effective Zeeman field}
The valley polarization induced by the magnetic proximity effect can be understood by a low-energy effective model\cite{ref8coupledspin,ref16Qi,ref17cheng},
for which the Hamiltonian at the K and K$^\prime$ valleys can be written as:
\begin{equation*}H=\frac{\Delta}{2} \hat{\sigma}_{z}-\lambda \tau \frac{\hat{\sigma}_{z}-1}{2} \hat{S}_{z}+\frac{\hat{\sigma}
_{z}-1}{2}\left(\hat{S}_{z}+\tau \alpha\right) B_v,
\end{equation*}
where $\Delta$ denotes the band gap, $\lambda$ the spin-orbit interaction strength, 
$\alpha_{Mo}$ the magnetic moment on Mo induced by the substrate and $B_v$ the effective Zeeman field for the valence band.
$\hat{\sigma}$ and $\hat{S}_{z}$ represent the Pauli matrix and the spin operator, respectively. 
$\tau$ is the valley index.
The effective Zeeman field for the valence band thus can be deduced by the valley polarization,
$\Delta E_{valley}=2\left(1+\alpha_{v}\right) B_{v}$.

\begin{table}
\centering
\caption{Effective Zeeman field for the magnetic proximity effect in MoTe$_2$/EuO.
Eu-octopolar, O-octopolar, O-P${(1\times2)}$ denote
the Eu-terminated octopolar-reconstructed, O-terminated octopolar-reconstructed,
and O-terminated P${(1\times2)}$ EuO(111) surfaces, respectively.
\textbf{\textit{S$_i$}} ($i=1,2,3$) denote the stacking configurations of MoTe$_2$/EuO. $\alpha_{\text{Mo} }$ denotes the induced magnetic momentum on Mo.
$E_{\text {valley}}$ defines the valley polarization for the valence band maximum. $B_v$ denotes the effective Zeeman field.}
\label{table1}
\begin{tabular}{lcccc}
\hline \hline Substrate & Configuration & $\alpha_{\text {Mo} }$ & $E_{\text {valley }}(meV)$ & $B_{\text {v} } (T) $ \\
\hline    & S1 & 0.023 & 4.67 & 40.34 \\
Eu-octopolar & S2 & 0.025 & 7.56 & 65.30 \\
          & S3 & 0.009 & 0.32 & 2.76 \\
\hline
          & S1 & 0.042 & 25.62 & 221.31 \\
O-octopolar & S2 & 0.069 & 43.43 & 375.15 \\
          & S3 & 0.017 & 3.54 & 30.58 \\
\hline
          & S1 & 0.024 & 12.07 & 104.26 \\
O-P${(1\times2)}$ & S2 & 0.019 & 32.77 & 283.07 \\
          & S3 & 0.019 & 38.00 & 328.24 \\
\hline
          & S4 & 0.078 & 3.17 & 27.38 \\
EuO(001) & S5 & 0.077 & 3.29 & 28.42 \\
           & S6 & 0.075 & 3.02 & 26.09 \\
\hline \hline
\end{tabular}
\end{table}

Table~\ref{table1} lists the parameters $\alpha_{Mo}$ and $\Delta E_{valley}$ obtained from DFT calculations 
and the estimated $B_{v}$ for MoTe$_2$/EuO.
For MoTe$_2$/EuO(111), $\alpha_{Mo}$ is dependent on the detailed interface structures, 
i.e., surface orientation and specific stacking configuration.
$B_v$ is determined by the induced magnetic momentum on Mo, i.e., $\alpha_{Mo}$.
Overall, $\alpha_{Mo}$ is larger for MoTe$_2$/EuO(001) than for MoTe$_2$/EuO(111).
Overall, $\alpha_{Mo}$ is larger for MoTe$_2$/EuO(001) than for MoTe$_2$/EuO(111).
Compared to those for MoTe$_2$/EuO(111), both $\alpha_{Mo}$ and $\Delta E_{valley}$ for MoTe$_2$/EuO(001) are insensitive to stacking.
The estimated values of $B_{v}$ for this type of heterostructure is about 27 T,
which is much smaller than those for MoTe$_2$/ideal-EuO(111),
but comparable to that for WSe$_2$/EuS(001)\cite{ref31zhao2017enhanced}.
These can be understood since EuO(001) and EuS(001) have the same surface atomic structure and 
the magnetic moments on the surface Eu atoms are almost the same for them.

In summary, we have investigated the structural and electronic structures of MoTe$_2$/EuO by means of first-principles calculations.
We have considered a number of surface reconstructions for the (111) surface of the substrate and 
find that they have substantial impact on the interface structure.
Consequently, the induced valley polarization in MoTe$_2$ by the reconstructed EuO(111) is at least one order of magnitude smaller than
that induced by the ideal EuO(111).
In addition, the surface reconstructions lead to that the Fermi levels of our systems are about 0.3 eV closer to the VBM than 
that of MoTe$_2$/ideal-EuO(111).
Moreover, we have also studied the valley polarization in MoTe$_2$/EuO(001).
For this type of heterostructure, our calculations find that the induced magnetic moments on the Mo atoms and valley polarization 
are almost the same for the considered stacking configurations.
We further estimate the effective Zeeman field induced by the substrate using a low-energy effective Hamiltonian.
The estimated effective Zeeman field for MoTe$_2$/EuO(001) is comparable to that for WSe$_2$/EuS(001).
Our study reveals that the magnetic proximity effect is strongly dependent on the substrate surface reconstruction and orientation,
which may be helpful in understanding recent experiments and designing interfaces for valleytronics. 

\begin{acknowledgments}
This work was supported by the National Natural Science Foundation of China (Grants No. 11774084, No. U19A2090, No. 51525202, and No. 11704117) and
the Project of Educational Commission of Hunan Province of China (Grant No. 18A003).
\end {acknowledgments}

\bibliography{references}
\bibliographystyle{apsrev4-1}

\end{document}